\documentclass[aps,prb,twocolumn,superscriptaddress]{revtex4-1}
\usepackage{amssymb}
\usepackage{physics}
\usepackage{graphicx}
\usepackage{amsmath}
\usepackage{amsthm}
\usepackage{amsfonts}
\usepackage[T1]{fontenc}
\usepackage{array}
\usepackage{multirow}
\usepackage{color}
\usepackage{esint}
\usepackage{bm}
\usepackage{color}
\usepackage{bbm}
\usepackage{hyperref}
\usepackage{babel}
\usepackage{titlesec}
\newcommand{\mean}[1]{\langle #1 \rangle}

\newcommand{\rr}{\bm{r}}

\begin{document}
	\title{Weak quantization of non-interacting topological Anderson insulator}
	
	\author{DinhDuy Vu}
	\author{Sankar Das Sarma}
	\affiliation{Condensed Matter Theory Center and Joint Quantum Institute, Department of Physics, University of Maryland, College Park, Maryland 20742, USA}
	
	\begin{abstract}
We study the transition between the two-dimensional topological insulator (TI) featuring quantized edge conductance and the trivial Anderson insulator (AI) induced by strong disorder. We discover a distinct scaling behavior of TI near the phase transition where the longitudinal conductance approaches the quantized value by a power law with system size, instead of an exponential law in clean TI.  This region is thus called the weak quantization topological insulator (WQTI). By using the self-consistent Born approximation, we associate the emergence of the weak quantization with the imaginary part of the effective self-energy acquiring a finite value at strong disorder. We use our analytical theory, supported by direct numerical simulations, to study the effect of disorder range on the topological Anderson insulator. Interestingly, while this phase is quite generic for uncorrelated or short-range disorder, it is strongly suppressed by long-range disorder, perhaps explaining why it has never been seen in solid state systems.
	\end{abstract}
	
	\maketitle
	\section{Introduction}
    Non-interacting topological phases are interesting not only from the theoretical perspectives but also for potential applications. Because of the nonlocal topological characteristics, the most important signature of these phases is the anomalous boundary metallic mode robust to perturbative deformations of the Hamiltonian. In this paper, we focus on two-dimensional (2D) topological systems featuring robust conducting edge modes \cite{Thouless1982,Kohmoto1985,Haldane1988,Hasan2010,Qi2011,Bernevig2013,Tudor2016}. A paradigm of this class, and probably the most experimentally viable, is the quantum spin Hall insulator induced by the spin-orbit coupling and band inversion \cite{Kane2005,Bernevig2006,Liu2008,Konig2007,Roth2009,Knez2011}. Such non-trivial topological insulators (TI) manifest an odd number of conducting helical modes at the boundary and a stable quantized conductance against small parameter changes. Recently, anomalous quantum Hall effect where chiral modes can exist without external field has also been reported \cite{Yu2010,Chang2013,Chang2016}. In all these experiments, quantized conductance is the key evidence to confirm the non-trivial topology.
    
    The intuition behind the stability of the anomalous edge states can be explained as follows. In a clean topological system, electrons at one edge can only move in one direction (per each time-reversal partner) so back scattering is absent and the current flows around the defect \cite{Buttiker1988}. As disorder becomes sufficiently strong, bulk parameters can be renormalized, driving the system to a different topological phase \cite{Shindou2009,Shinsei2012,Groth2009,Li2009,Jiang2009, Fu2012,Xu2012,Song2012,Song2014,Qin2016,Liu2017,Huang2018,Krishtopenko2020,Hua2021,Haim2019}. An interesting case is when the renormalized Hamiltonian itself supports non-trivial topology with robust boundary modes while the original clean system is trivial \cite{Groth2009,Li2009,Jiang2009,Li2011,Xu2012,Song2012,Song2014,Qin2016,Haim2019,Krishtopenko2020,Liu2020,Hua2021}, the so-called topological Anderson insulator (TAI) phase. This phenomenon can also be studied in one-dimensional systems \cite{Altland2014,Mondragon2014,Velury2021,Lin2021} and has been observed experimentally in atomic and optical systems \cite{Meier2018}, even though the topology here is characterized through a bulk index rather than the associated gapless boundary mode. The quantization plateau should not survive to an arbitrarily large disorder strength, with the bulk eventually becoming the exponentially localized Anderson insulator (AI) with trivial topology and thus zero edge conductance. While early works numerically demonstrated  a ``levitation and pair annihilation'' mechanism for the suppression of edge conductance \cite{Onoda2003,Onoda2007}, an intuitive physical picture is provided through a percolation process. Conducting bands, in the presence of disorder, generically develop tails of exponentially localized states that eventually overlap with the bulk gap \cite{Xu2012,Zhang2013}. These localized  bulk ``islands'' become connected if the disorder landscape is correlated and form a percolating network, effectively acting as a passage for the two edge modes to percolate into the bulk and destroy the quantization plateau through an effective edge-edge coupling \cite{Girschik2013,Girschik2015}.   In comparison with the pristine system, the bulk density of state at the energy gap is exactly zero, so the edge-edge hybridization can only happen through tunneling across the vacuum bulk, which is exponentially suppressed with the distance. 
 
    In this paper, we study the transition regime between  TI and AI with increasing disorder strength, focusing on the quantization plateau (and its eventual disappearance) in the two-terminal longitudinal conductance. We first numerically show that near the TI-AI phase boundary, the scaling of the conductance quantization error $(1-G)$ with the system size is slower than any exponential laws usually observed in the clean TI. We refer to this region as the weak quantization topological insulator (WQTI), while reserve the TI nomenclature for the regime with the exponentially fast quantization. This change of the scaling law distinguishes TI and WQTI. In the following, we show that this transition is the manifestation of the second-order transition of the imaginary part of the self-energy acquiring non-zero value for sufficiently strong disorder [see Fig.~\ref{fig2} and Fig.~2(a) of \cite{Supplement}].
    
    Here, we describe how the WQTI fits into other phase classification schemes in the present literature. In terms of topological classification, WQTI and TI belong to the same phase and should exhibit perfect quantized conductance in the thermodynamic limit. However, for finite systems, TI (WQTI) would manifest robust (fragile) quantization,  which has implications for experiments. This motivates our separation of WQTI from the TI phase based on the scaling behavior. We note that TAI refers to the disorder-induced quantized conductance phase starting from the pristine limit with no metallic edge modes, either because the bare (before the disorder-induced renormalization) mass is positive or the bare chemical potential lies outside the bulk gap. Therefore, TAI may overlap with either TI or WQTI depending on details. 
    
    A key distinguishing feature of our work compared with all earlier works reporting similar numerical evidences is the development of an analytical framework that theoretically establishes the fragile power-law behavior of the WQTI phase. Furthermore, we predict the emergence of TI, WQTI and AI phases based entirely on our analytical theory, obtaining excellent agreement with the exact numerical results. To demonstrate the predictive power of our theory, we study the effect of disorder range  on the TAI region (with no quantized-conductance clean analog). Both our analytic theory and numerical simulation show a progressive suppression of the quantization plateau as the disorder range increases. Again, this has been reported based on numerical simulations before \cite{Girschik2013,Girschik2015}, but our analytical model provides deep insight into the underlying physics. The main text of this Letter is devoted to the onsite disorder so, with no loss of generality, we only work on one spin sector of the Hamiltonian, leaving the other time-reversal partner implicit.
 
     \begin{figure}
 	\centering
 	\includegraphics[width=0.43\textwidth]{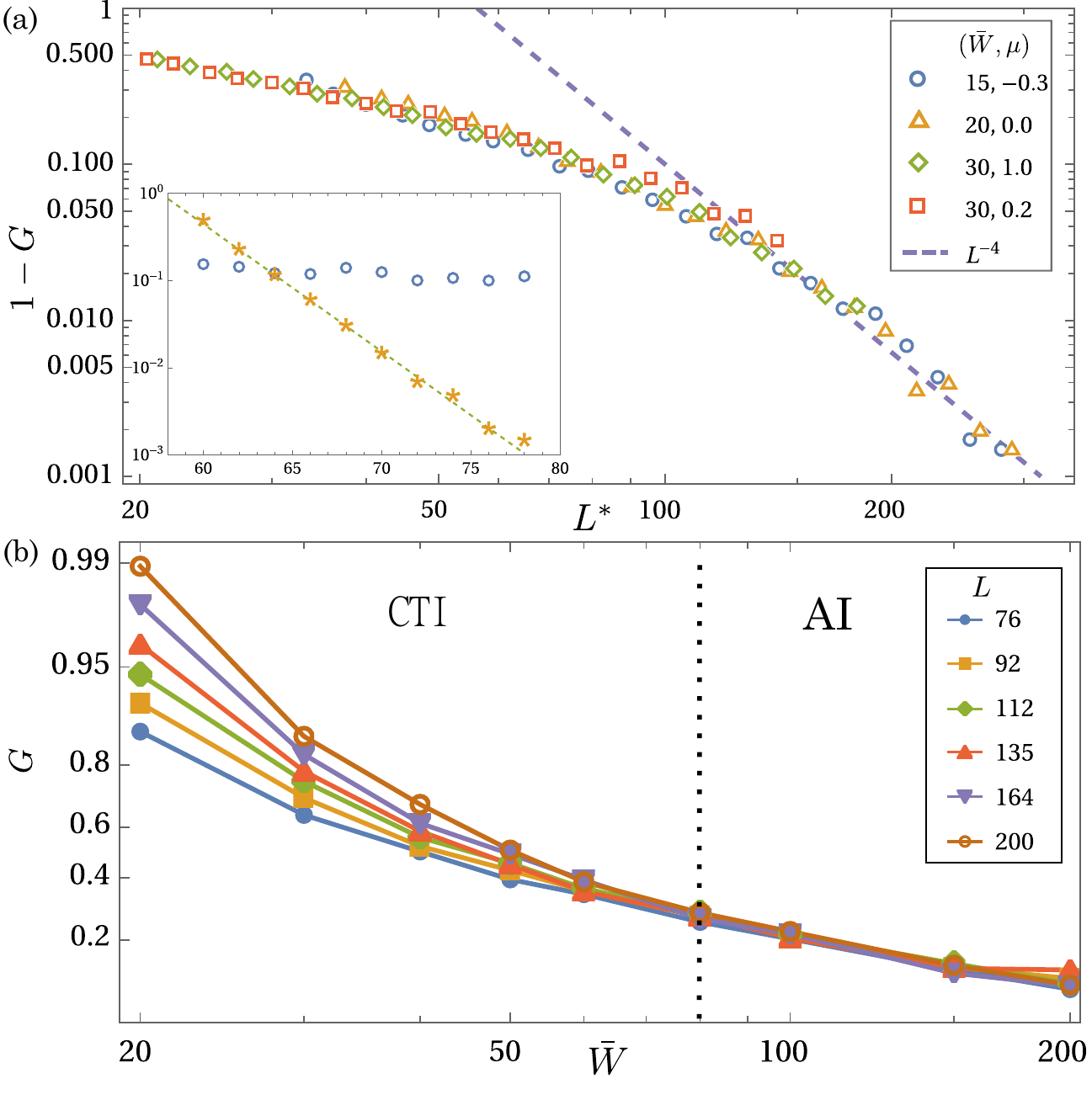}
 	\caption{(a) The quantization error $1-G$ with the rescaled size $L^*$ so that $1-G$ at different parameters collapses into a single function with an asymptote $L^{-4}$. The inset shows the finite-size scaling of quantization error but in a log-linear graph, distinguishing the exponential decay for $W=10,\mu=0.5$ (yellow asterisks) from the polynomial decay for $W=15,\mu=-0.3$ (blue circles). (b) The scaling of the conductance at $\mu=0$ across the WQTI-AI transition. The vertical axis is plotted by the scale $\ln[y/(1-y)]$ to show both the limits near zero and one. The lattice parameters for the simulation are identical to Fig.~\ref{fig2}(b)\label{fig1}. }
 \end{figure} 
    \section{Theory}
    We start with a clean Chern insulator Hamiltonian  that preservers the translational symmetry
    \begin{equation}\label{Hamiltonian}
    	H_0(k_x,k_y) = \alpha(k_x \sigma_x-k_y\sigma_y) + (m+\beta k^2)\sigma_z +\gamma k^2 \sigma_0.
    \end{equation}
    The tunable chemical potential is $\mu$. We perform the numerical simulation on a square lattice with the lattice constant $a$, using a discretized version of Eq.~\eqref{Hamiltonian}. The parameters $\alpha, \beta,\gamma$ when accompanied by an appropriate power of the lattice constant give the unit of energy, i.e. $\alpha/a$, $\beta/a^2$, and $\gamma/a^2$ have the same dimension as the mass $m$.  Therefore, throughout this paper, we fix $a=1$ and provide the value for $\alpha,\beta,\gamma$ with the length scale $a$ implied. To reproduce the effect of quenched impurities, we introduce random disorder at each site whose strength is chosen independently from a uniform distribution $[-W/2, W/2]$. Each impurity interacts with electrons via a Gaussian interaction, resulting in the disorder potential being long-range correlated $\mean{u(\rr)u(\bm{r}')}\sim e^{-(\rr-\rr')^2/2\xi^2}$ \cite{Supplement} ($\mean{\cdot}$ denotes averaging over disorder configurations.) The disordered Hamiltonian is thus $H_0+u(\rr)\sigma_0$ with $\sigma_0$ reflecting the on-site disorder (other types of disorder differing by the accompanying matrix are studied in the Supplemental Materials \cite{Supplement}). Because of the finite correlation length, $W,\alpha$ are rescaled by $\xi^{-1}$ and $\beta,\gamma$ by $\xi^{-2}$ \cite{Supplement}. For this reason, we use the rescaled disorder strength $\bar{W}=W/\xi$ instead of the absolute value in all the figures. The longitudinal conductance $G$ is computed exactly using the Landauer-B\"{u}ttiker formalism \cite{Landauer1970,Buttiker1988} and implemented numerically using the Kwant package \cite{Groth2014}. We average over up to 700 disorder configurations to ensure convergence and present the conductance in the unit of $e^2/h$ so that the ideal TI quantized conductance is unity.
    
\begin{figure}
	\centering
	\includegraphics[width=0.45\textwidth]{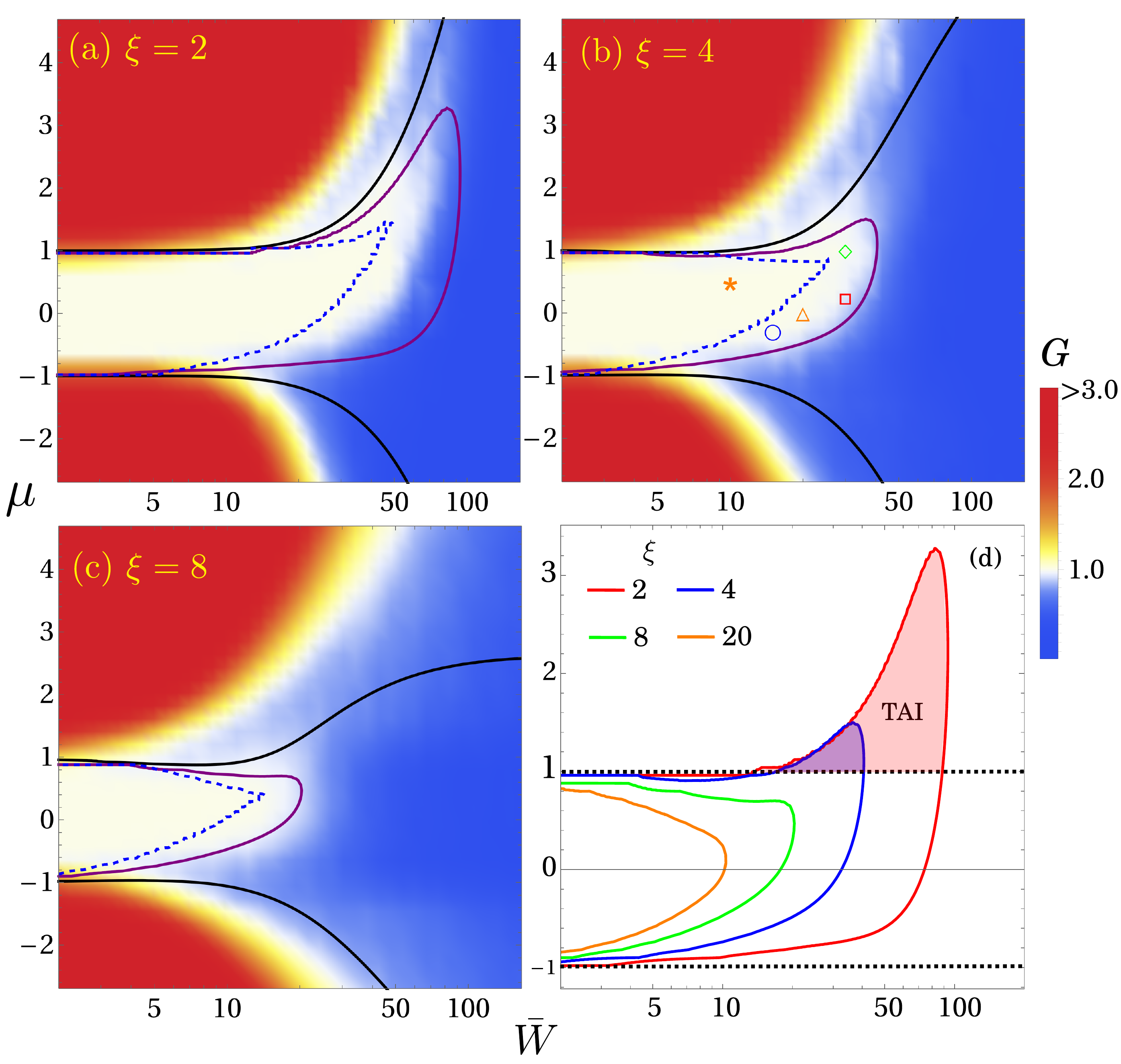}
	\caption{Longitudinal conductance with increasing disorder correlation lengths (a-c). For the specifications of the lattice, the width is $200a$, the length is $400a$, $m=-1, \alpha=16, \beta=100,\gamma=48$. Points marked in panel (b) correspond to Fig.~\ref{fig1}(a). The black lines show the renormalized energy gap $\bar{\mu}=\pm\bar{m}$, the purple lines are the $\Xi=2$ isolines in Eq.~\eqref{closed_bound}, and the dashed blue lines are the boundary of the $\Im\Sigma_0=0$ region. (d) The $\Xi=2$ isolines (solid lines) at different correlation lengths. The dotted lines mark the energy gap in the pristine limit. Accordingly, the TAI phase (shaded region beyond the dotted lines) only exists for $\xi=2$ and $4$. \label{fig2} }
    \end{figure}

    From the theoretical perspective, we can average the Green function over disorder configurations to obtain an effective theory with recovered translational symmetry, i.e. $G=\mean{\left[\mu-H_0-u(\rr)\sigma_0\right]^{-1}} = (\mu-H_0-\Sigma)^{-1}$.
    This self-energy $\Sigma$ is a matrix $\Sigma=\Sigma_0\sigma_0 + \Sigma_z\sigma_z$ (by symmetry reason there are no $\sigma_{x,y}$ terms). If we only consider non-crossing diagrams (i.e. the self-consistent Born approximation), the self-energy can be written as an integral equation \cite{Supplement}. Due to renormalization by the self-energy, the renormalized mass $\bar{m} = m+\text{Re}\Sigma_z$ and chemical potential $\bar{\mu} = \mu - \text{Re}\Sigma_0$ define a new energy gap by the condition $\bar{m}<0$ and $|\bar{\mu}| < |\bar{m}|$. However, as shown in the numerical results (Fig.~\ref{fig2}), the boundary generated from the gap-opening condition, defined only by the real parts of the self-energy, does not enclose the TI region but in fact extends far beyond. Thus, a theory based only on mass and chemical potential renormalization, described by the real part of the self-energy is incomplete. We take into account the imaginary part of the self-energy analytically and obtain the boundary enclosing the numerically simulated quantization plateau, thus providing a complete and correct theory.

    We first provide a preliminary argument. The gapless boundary modes inherit the imaginary term from the bulk which can be regarded as the incoherent broadening of the edge excitation. As such, the edge current can disperse into the bulk where it might hybridize with the current leaking from the opposite edge and mutually exchange momentum. This effective coupling between the two chiral edges, arising from the impurity-scattering-induced imaginary self-energy, leads to the quantization error and eventually the suppression of the conducting edge modes. This is the physical mechanism driving the TI-WQTI-AI transition with increasing charge disorder. Formally, when the two edge states hybridize, the quantization error in the longitudinal conductance is proportional to hybridization probability. In the Supplemental Materials \cite{Supplement}, we evaluate this probability as $\sim F(\Delta E/\Gamma)L^{-4}$. Here, $F$ is a function, $\Delta E$ is the energy separation between the renormalized chemical potential and the edges of the renormalized bulk gap, $\Gamma$ is the energy level broadening and $L$ is the inter-edge separation.  The level broadening may arise from either $\Im\Sigma_0$ or $\Im\Sigma_z$, but the edge states are the eigenvectors of $\sigma_x$, so only the former can contribute to the $L^{-4}$ scaling.
    We thus identify $\Gamma=\Im\Sigma_0$. From this dimensional scaling analysis, if the imaginary part of the self-energy is non-zero, $1-G$ should scale as $L^{-4}$; and, if it is zero, the quantization converges exponentially because the two edges can only interact through tunneling which is suppressed exponentially in the absence of level broadening.
 
    To quantify this physics, we numerically compute the quantization error $1-G$ while changing the system width (the ratio length/width is fixed at $2$). For points with non-zero $\text{Im}\Sigma_0$ [see Fig.~\ref{fig2}(b)], $1-G$ decreases with increasing system size by a sub-exponential scaling law as shown in the log-log Fig.~\ref{fig1}(a). After rescaling $L$, there emerges a one-parameter scaling function $\beta=d\ln(1-G)/d\ln L$ that approaches $-4$ for $G$ sufficiently close to unity, consistent with our analytical picture of the inter-edge hybridization through the bulk leakage. Because of this power-law scaling of the quantization error, we refer to this phase as the WQTI. Compared with the pristine TI, we compute the quantization error for a point with vanishing $\text{Im}\Sigma_0$ (but a finite strength of disorder) which, as shown in the inset of Fig.~\ref{fig1}(a), has a clear exponential scaling. This comparison establishes that the disorder-induced TI-WQTI transition is driven by the imaginary part of the 
    self-energy acquiring a finite value beyond a critical point, resulting in the edge localization length diverging across the phase transition (and thus hybridizing in the WQTI phase). While earlier works numerically demonstrated the strong finite-size effect observed in the presence of strong disorder \cite{Li2011,Chen2012,Xu2012}, we quantify this behavior by providing the one-parameter scaling function and an analytical and physical explanation which remarkably reproduces the asymptotic scaling exponent of $-4$.  
     
    Now, we discuss what happens if the disorder increases further. In Fig~\ref{fig1}(b), we increase the disorder strength and study the WQTI-AI transition. For $G\gtrsim0.3$, the scaling flow is to $1$ with the rate being progressively faster for larger (closer to $1$) $G$. On the contrary, when the disorder is strong enough, the conductance is suppressed close to zero, and more importantly, does not depend on the system size. This change of scaling marks the WQTI-AI phase transition. The small finite conductance in the AI phase might be caused by rare conducting bubbles. These bubbles are incoherent, so the finite-size scaling law should obey the Ohmic law, i.e. $L^0$ for 2D, albeit the conductance magnitude is much smaller than the conductance quantum because of strong localization with $G\sim 0$. Combining with the previous argument, the scaling exponent $\beta$ is negative for $G>G_c\sim 0.3$ and approaches $-4$ in the limit $G\to 1$; for $G< G_c$, $\beta=0$ \cite{Supplement}. We note one difference with the transverse conductance measurement in which $\beta <0 (>0)$ for $G>0.5 (<0.5)$ and there is only one fixed point ($\beta=0$) at $G=0.5$ \cite{Onoda2003}. The reason for this disparity is that the rare conducting bubbles support longitudinal but not transverse conductivity.  
     
     While the TI-WQTI boundary can be obtained analytically from $\text{Im}\Sigma_0$, it is not so obvious for the WQTI-AI transition because of very strong disorder. Instead, we look at where our self-energy approximation breaks down and the Anderson localization physics prevails. The hybridization probability already shows that the factor $\Gamma/\Delta E$, if being large, can significantly couple the two edges and can completely suppress the edge conductance. Since the hybridization can happen through either the upper or lower bulk bands, we propose a quantity to capture its magnitude
     \begin{equation}\label{closed_bound}
       \Xi = \frac{\Im\Sigma_0}{\bar{\mu}+|\bar{m}|} + \frac{\Im\Sigma_0}{|\bar{m}|-\bar{\mu}}.
     \end{equation}   
     The physical driving mechanism of the WQTI-AI transition is the disorder-induced density of percolating states inside the energy gap, which can be estimated by $\rho(\mu)\sim \text{Im}\Sigma_0/\left(\bar{m}^2-\bar{\mu}^2\right)$. This shows that Eq.~\eqref{closed_bound} is indeed consistent with the microscopic physics. Therefore, we expect the $\Xi-$function including the imaginary part of the self-energy to resolve the inconsistency between the renormalized energy gap (obtained from only the real part of the self-energy) and the numerically generated conductance. In particular, by fixing $\Xi$ to a constant, we can reproduce the boundary of the numerical quantization regime. To verify this, we apply our theory to explain the dependence of the TAI phase on the correlation length of the disorder. This effect has been studied in a few papers \cite{Girschik2013,Girschik2015}, based mostly on numerical simulations.
     
     \section{Numerical phase diagram}
     \subsection{$m<0$} 
     We first study the case where the clean limit is a non-trivial TI, i.e. for $W=0$, the longitudinal conductance is quantized for $m<\mu<-m$. To connect with our finite-size simulation, we identify the region of $G$ within $10\%$ error from unity as the quantization plateau, rather than the critical value $0.3$ derived earlier from the scaling analysis. The first remark is that while the renormalized energy gaps (black lines) are clearly larger than the quantization regime, the isolines $\Xi=2$ (purple lines) consistently enclose this region, even as $\xi$ changes. These results also agree with the absence of  quantization plateau in the $\mu<m$ region, despite still being within the energy gap. This is a strong validation of the theory.
     
     We now focus on the fate of the TAI in the presence of long-range correlated disorder. Although a prominent TAI region, i.e. the part of the quantization plateau with $\mu>-m$, is visible for $\xi=2$ [Fig.~\ref{fig2}(a)], it is suppressed quickly with longer correlation lengths [Fig.~\ref{fig2}(b) and (c)]. The trend appears more apparent in Fig.~\ref{fig2}(d) where we collect all the $\Xi-$isolines for different correlation lengths. The naive explanation is that the exponential suppression induced by $\xi$ prevents the energy gap to broaden \cite{Supplement}, thus ruling out the TAI. This limit is approximately reached for $\xi=20$ indicated by the almost symmetry around $\mu=0$.  However, the TAI already vanishes for $\xi=8$ despite a visibly widened energy gap. In fact, we show that the $\xi-$dependence is mostly due to the rapid increase of $\text{Im}\Sigma_0$ passing the critical point, rather than the decrease of the gap \cite{Supplement}.

     \begin{figure}
	\centering
	\includegraphics[width=0.45\textwidth]{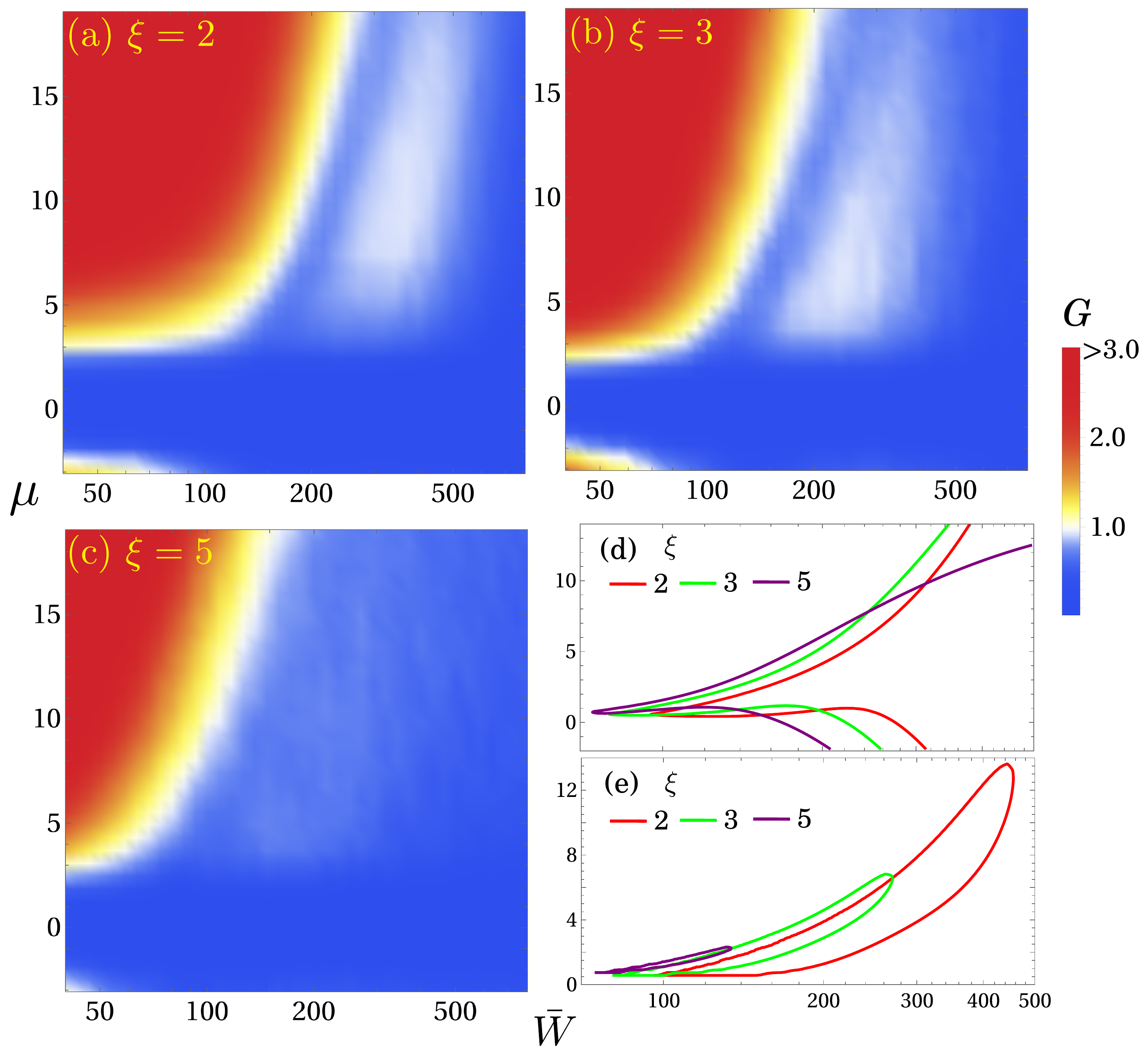}
	\caption{(a-c) Numerical simulations for the trivial clean limit. The width is $350a$, the length is $700a$, $m=1, \alpha=120, \beta=400,\gamma=200$. (d) The renormalized energy gaps for different disorder correlation lengths corresponding to (a-c). (e) Similar to (d) but for the $\Xi=2$ isolines. \label{fig3}}
     \end{figure}
     
     \subsection{$m>0$}
     We now consider the case where disorder inverts the sign of the mass, thus introducing non-trivial topology to an otherwise trivial system. By definition, the whole quantization plateau is now TAI. As can be seen from Figs.~\ref{fig3}(a)-(c), the TAI phase is suppressed for long-ranged disorder, similar to the case of $m<0$. Again, the naive argument predicts that the TAI should disappear when a sufficiently long correlation length suppresses the mass inversion by rendering the quadratic terms in the Hamiltonian irrelevant  \cite{Groth2009,Li2009,Krishtopenko2020}. However, our numerical simulations show a complete loss of quantization even at a moderate $\xi=5$ where the mass inversion is still clearly present. On the other hand, the isoline $\Xi=2$ exhibits a consistent shrinkage of the TAI phase. We note that the quantitative agreement here between the theoretical and numerical results is not as good as the $m<0$ case. The reason might be that the mass first needs to be inverted so the TAI plateau now exist at a very high disorder strength. Our analytical theory only includes non-crossing diagrams \cite{Supplement}, so we expect its accuracy to degrade at very strong disorder. Nevertheless, theoretical and numerical results in Fig.~\ref{fig3} show similar trends. These two examples demonstrate that our theory, particularly the $\Xi-$function, is an effective tool to study the disorder-driven topology.
     
     \section{Conclusion}
     We theoretically identify several different class A disordered 2D TI phases, including TI (perturbatively weak disorder), WQTI (moderate disorder), and AI (strong disorder). Our analytical theory, directly supported by numerical simulations, relies on the imaginary part of the disorder-induced self-energy, which was ignored in earlier studies. One might raise a question about the extremely accurate quantization in integer quantum Hall experiments. In addition to various quantitative reasons, e.g. large gap, a protective mechanism is based on the bulk of a Hall insulator being localized at essentially all energy. This is because the Landau levels are exactly flat, and in the presence of disorder, quickly collapse into localized states (except when the sample is very small). The WQTI mechanism is not possible because it requires extended bands around the energy gap. 
     
     In earlier works \cite{Li2009,Groth2009}, the disorder is uncorrelated and zero-range, and the TAI is predicted to be a generic phase. However, our analysis with non-zero correlation lengths shows that TAI is fragile and vanishes for long-range disorders. In 2D semiconductors \cite{Kane2005,Bernevig2006,Liu2008}, the dominant disorder is always the long-range correlated Coulomb disorder arising from random quenched charge impurities \cite{Watson2011,DasSarma2015}, and hence the TAI phase is difficult to observe (except perhaps in artificial AMO systems \cite{Meier2018,Liu2020}).  Our work establishes TI, WQTI, and AI as the three generic phases in the presence of disorder, with WQTI being a somewhat fragile intermediate critical phase in between the weak-disorder TI and the strong-disorder AI phase.
     
     Lastly, we compare our ``weak quantization'' with the ``quantization loss'' due to intra-edge coupling between the two spin sectors in quantum spin Hall insulators \cite{Schmidt2012,Vayrynen2013}. Being a local process, the latter cause finite deviation from the quantized conductance even in the thermodynamic while in our model the quantization recovers slowly. Moreover, the intra-edge backscattering is not possible in non-interacting systems.
    
     \begin{acknowledgments}
     		\textit{Acknowledgments - }
     		This work is supported by and Laboratory for Physical Sciences. This work is also generously supported by the High Performance Computing Center (HPCC) at the University of Maryland.
     \end{acknowledgments}

     \bibliographystyle{apsrev4-1}
     \bibliography{TAI_bib}

\end{document}


\title{Supplemental Material for ``Weak quantization of non-interacting topological Anderson insulator''}
\author{DinhDuy Vu}
\author{Sankar Das Sarma}
\affiliation{Condensed Matter Theory Center and Joint Quantum Institute, Department of Physics, University of Maryland, College Park, Maryland 20742, USA}
\maketitle

\section{Long-range correlated disorder and self-consistent Born approximation}
\subsection{Long-range interacting impurity}
On a square lattice, at each site $\bm{R}$, we introduce an impurity whose strength $W(\bm{R})$ independently takes a random value from a uniform distribution $[-W/2,W/2]$. These impurities are uncorrelated by construction
\begin{equation}
	\mean{W(\bm{R})} = 0,\quad \mean{W(\bm{R})W(\bm{R}')} = \frac{W^2}{12}\delta_{\bm{R},\bm{R}'} = \frac{W^2a^2}{12}\delta(\bm{R}-\bm{R}')
\end{equation}
The lattice constant $a$ appears when we promote the discrete Kronecker to the continuum field description because we place an impurity every distance $a$ in a two dimensional lattice. We now argue that if these impurities interact with electrons via a finite-range interaction, the effective disorder field felt by an electron is effectively long-range correlated. We assume this electron-impurity interaction is given by a Gaussian function so that the net disorder potential is 
\begin{equation}
	u(\rr) = \int  \frac{e^{-(\rr-\bm{R})^2/\xi^2}}{\pi \xi^2}W(\bm{R})d^2\bm{R}.
\end{equation}
As a result, $u(\rr)$ is spatially correlated by
\begin{equation}\label{correlated}
	\begin{split}
	 \mean{u(\rr)u(\rr')} = \frac{W^2a^2}{12\pi^2\xi^4}\int e^{-(\rr-\bm{R})^2/\xi^2}e^{-(\rr'-\bm{R})^2/\xi^2}d^2\bm{R} = \frac{W^2a^2}{24\pi\xi^2}e^{-(\rr-\rr')^2/2\xi^2}
	\end{split}
\end{equation}

\begin{figure}
	\centering
	\includegraphics[width=0.95\textwidth]{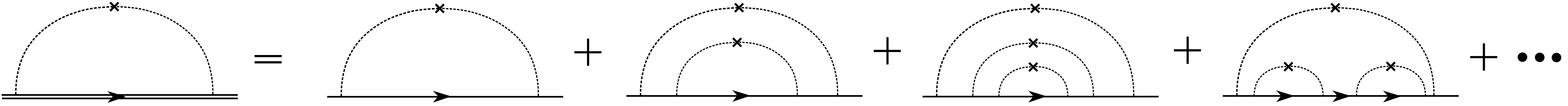}
	\caption{Self-consistent diagram for the disorder-induced self-energy. The double-line represents the renormalized Green function with the self-energy included. The cross denotes the impurity and the dashed lines denote the impurity-electron interaction. The self-consistent diagram can be thought of as the sum of all non-crossing diagrams with single solid lines representing the bare Green function. After averaging over disorder configurations, each dashed line segment connecting two points of a Green function is substituted by Eq.~\ref{correlated}. \label{fig1}}
\end{figure}

The self-consistent Born approximation sums all the non-crossing diagrams as shown in in Fig.~\ref{fig1}. We also do not consider more-than-two-point correlation functions as in the $T-$matrix approximation.  After averaging over disorder configurations, the theory recovers the translational symmetry so we can compute the self-energy self-consistently as
    \begin{equation}\label{selfconsistent}
\begin{split}
\Sigma &= \frac{W^2a^2}{12}\int \frac{d^2\kk}{(2\pi)^2} \frac{e^{-k^2\xi^2/2}}{\mu-H(m,\alpha,\beta,\gamma;\kk)-\Sigma} = \frac{W^2a^2}{12\xi^2}\int \frac{d^2\kk}{(2\pi)^2} \frac{e^{-k^2/2}}{\mu-H(m,\alpha/\xi,\beta/\xi^2,\gamma/\xi^2;\kk)-\Sigma}
\end{split}
\end{equation}
The second equality is constructed by rescaling the integration variable $\kk\to \kk\xi$. Compared with the case of uncorrelated disorder, the extra exponential decay comes from the Fourier transform of the finite-range function $O(\rr-\rr')=\mean{W(\rr)W(\rr')}$. This term provides a natural cutoff for the integral so we do not need to impose an artificial UV cutoff. For brevity, we define the rescaled quantities $\bar{W}=W/\xi, \bar{\alpha}=\alpha/\xi, \bar{\beta}=\beta/\xi^2$, and $\bar{\gamma}=\gamma/\xi^2$. The integral in Eq.~\ref{selfconsistent} can be performed exactly, yielding an explicit systems of equation
\begin{equation}\label{SCsystem}
\begin{split}
 &\Sigma_0 = \frac{\bar{W}^2}{48\pi \Delta}\left[ F(x_2)\left(\mu-\Sigma_0 + 2\bar{\gamma}x_2\right) - F(x_1)\left(\mu-\Sigma_0 + 2\bar{\gamma}x_1\right)\right],\\
 &\Sigma_z = \frac{\bar{W}^2}{48\pi \Delta}\left[ F(x_2)\left(m+\Sigma_z - 2\bar{\beta}x_2\right) - F(x_1)\left(m+\Sigma_z - 2\bar{\beta}x_1\right)\right]
\end{split}
\end{equation}
Here, we decompose the self-energy matrix as $\Sigma=\Sigma_0\sigma_0+\Sigma_z\sigma_z$. Other terms in Eq.~\ref{SCsystem} are given by
\begin{equation}
\begin{split}
  & F(x) = e^x\left[i\pi \text{sign}[\text{Im}(x)] + \text{Ei}(-x)\right]\\
  &\Delta = (a^2-4bc)^{1/2},\quad x_{1,2} = \frac{a\mp \Delta}{4b},\\
  &a=-\bar{\alpha}^2-2\bar{\gamma}(\mu-\Sigma_0)-2\bar{\beta}(m+\Sigma_z),\quad b=\bar{\gamma}^2-\bar{\beta}^2, \quad c=(\mu-\Sigma_0)^2-(m+\Sigma_z)^2
\end{split}
\end{equation}
The solution of the Eq.~\eqref{SCsystem} is transcendental but can be obtained quickly through multiple numerical iterations. Then, the renormalized mass term and chemical potential are
\begin{equation}
	\bar{\mu} = \mu -\text{Re}(\Sigma_0),\quad \bar{m} = m +\text{Re}(\Sigma_z).
\end{equation}
We emphasize that it is necessary to obtain the full self-consistent solution. If one assume $\Sigma=0$ in the RHS of Eq.~\ref{selfconsistent}, a simple expression for the solution can be found but its accuracy might be limited and more importantly, it is not clear how the imaginary part of the self-energy can be obtained.

\subsection{$L^{-4}$ scaling of the WQTI phase}
In this subsection, we show that quantized conductance is reduced by the edge-edge hybridization probability, which is suppressed by $L^{-4}$ in the weak-coupling limit. The longitudinal conductance is given by $G=e^2v_F\rho$ where $v_F$ is the Fermi velocity and $\rho$ is the density of states at the Fermi energy. For a chiral state $\psi(x) = D^{-1/2}e^{ik_Fx}$ with $D$ being the longitudinal dimension for normalization, $v_F=\text{Re}\bra{\psi}i\partial_x\ket{\psi}/m_e$ and $\rho=1/2\pi\hbar v_F$, leading to the quantized conductance $G=e^2/h$. We assume that a small coupling between opposite chiral edges, which effectively acts as a backscattering, mainly reduces the propagating velocity. Specifically, if $\psi(x)\to\psi'(x)\sim \sqrt{1-|\eta|^2}e^{ikx} + \eta e^{-ikx}$ with $\eta$ being the mixing amplitude, then $v_F\to v_F' = v_F(1-2|\eta|^2)$. As a result, the conductance loss $1-G$ is proportional to the hybridization probability.

We now estimate the scaling of the hybridization probability. We assume that the probability of an excitation of energy $E$ to overlap with an eigenstate of energy $E+\Delta E$ is given by a function $F(\Delta E /\Gamma)$ where $\Gamma$ is the self-energy-induced incoherent level broadening. Here, $F(x)$ decreases with increasing $|x|$. Eigenstates within the energy band gap  decay exponentially into the bulk, so the main contribution comes from states near the gap edge where the localization length diverges. Specifically, the probability of an edge mode of energy $E_0$ to appear at a distance $y$ in the bulk is
\begin{equation}\label{scaling}
P(E_0, y) = \int F\left(\frac{E-\bar{E}_0}{\Gamma} \right) e^{-Cy\sqrt{|\bar{m}|-E}}dE \propto y^{-2}F\left(\frac{|\bar{m}|-\bar{E}_0}{\Gamma} \right).
\end{equation}
Here, $\bar{E}_0$ is the disorder-renormalized $E_0$ and $|\bar{m}|$ is the upper edge of bulk gap with $\bar{m}$ being the renormalized mass in the disorder-averaged Hamiltonian. The same argument can be applied to the leakage into the bulk through the lower edge $-|\bar{m}|$ of the bulk band.  The energy level broadening is $\Gamma=\text{Im}\Sigma_0$. The probability of inter-edge overlap then follows analytically  from Eq.~\eqref{scaling} to scale as $L^{-4}$, where $L$ is the sample width or the inter-edge separation. We compare Eq.~\ref{scaling} with the disorder-induced density of states inside the bulk gap
\begin{equation}
	\rho(E_0) = \frac{1}{\pi}\left[ \text{Tr} \int\frac{d^2\kk}{E_0-H_0-\Sigma}  \right] \approx -\frac{1}{\pi} \int\frac{ \text{Tr}\left[(\text{Im}\Sigma_0)\sigma_0 +(\text{Im}\Sigma_z)\sigma_z\right]}{(\bar{m}^2-\bar{E}_0^2) + \alpha^2k^2 +\mathcal{O}(k^4)}   d^2\kk \sim \frac{\text{Im}\Sigma_0}{\bar{m}^2-\bar{E}_0^2}.
\end{equation}
As $\rho(E_0)$ increases, the edge-edge hybridization probability grows concomitantly. This shows the consistence of our Born approximation with the microscopic picture that in the WQTI phase, the two edges couple through bulk percolating states inside the bulk gap.

\subsection{TAI suppression by long-range disorder}

Even though an apparent solution for the self-energy is generally not known, some properties can be derived in the large-$\xi$ limit where $\bar{\beta},\bar{\gamma}$ and $\bar{\alpha}^2$ are suppressed by a factor $\xi^{-2}$. We can then simply Eq.~\eqref{SCsystem} into
\begin{equation}
 \begin{split}
  &\Sigma_0 = \frac{\bar{W}^2}{24\pi} \frac{\mu-\Sigma_0}{(\mu-\Sigma_0)^2-(m+\Sigma_3)^2} \\
   &\Sigma_z = \frac{\bar{W}^2}{24\pi} \frac{m+\Sigma_z}{(\mu-\Sigma_0)^2-(m+\Sigma_z)^2} 
 \end{split}
\end{equation}
Then it easy to show that if one start at the clean energy gap $\mu=\pm m$ (actually $|\mu|=|m|-\epsilon$ with infinitesimal $\epsilon$ to avoid singularity) then the renormalized value satisfies $\bar{\mu}=\pm\bar{m}$, i.e. the width of the bulk gap is unchanged by the disorder, ruling out the existence of the TAI phase. In addition, one has the symmetry $\Sigma_0\to-\Sigma_0$ for $\mu\to-\mu$, so the the solution is symmetric around $\mu=0$.  This limit is approximately reached when $\xi^2\sim (\gamma,\beta,\alpha^2)/|m|$. However, we now argue the TAI phase is suppressed much sooner than this limit mainly due to the fast growth of the imaginary part of the self-energy.

\begin{figure}
	\centering
	\includegraphics[width=0.95\textwidth]{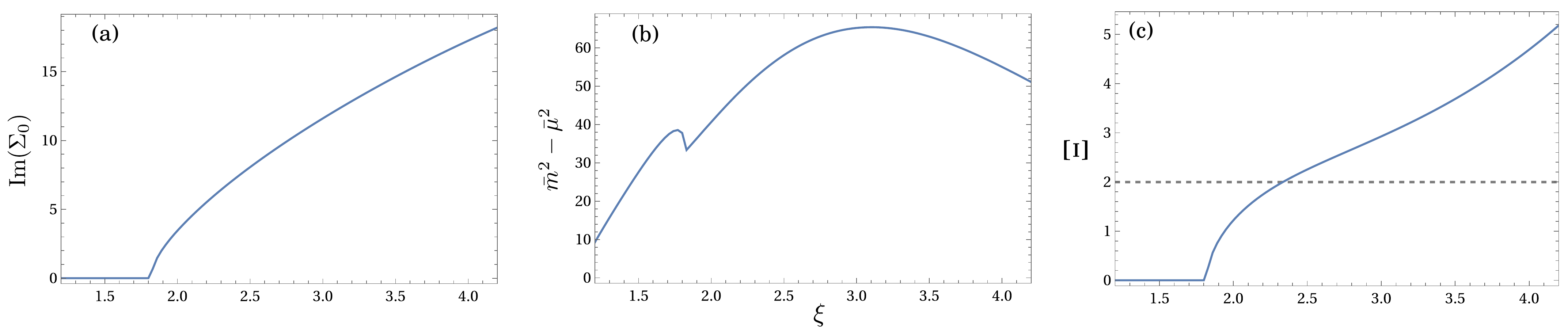}
\caption{Variations of $\text{Im}(\Sigma_0)$ (a), $\bar{m}^2-\bar{\mu}^2$ or the squared gap width (b), and $\Xi$ (c) as a function of the disorder correlation length $\xi$. The parameters are $m=1, \alpha=120,\beta=400,\gamma=200,\bar{W}=300,\mu=5$ (corresponding to Fig.3 in the main text). \label{fig2}}
\end{figure}

As shown in Fig.~\ref{fig2}(a), the imaginary part of the Green function grows rapidly beyond the critical point. Comparing  Figs.~\ref{fig2}(a) and (c), the variation of the $\Xi$-factor used to bound the TAI region is almost concomitant with that of the imaginary part. As a result, the threshold $\Xi=2$ is reached when the renormalized gap is still open and $\bar{m}<0$. For the given parameters, based on the gap-opening condition only, the threshold $\xi$ for a complete TAI loss should be 8; while according to our numerical simulation, this point $\bar{W}=300,\mu=5$ is already AI for $\xi=3$, which agrees better with the threshold $\xi=2.5$ from the self-consistent Born approximation [see the crossing point in Fig.~\ref{fig2}(c)].

\section{Additional numerical data}

\subsection{One-parameter scaling function}
\begin{figure}
	\centering
	\includegraphics[width=0.9\textwidth]{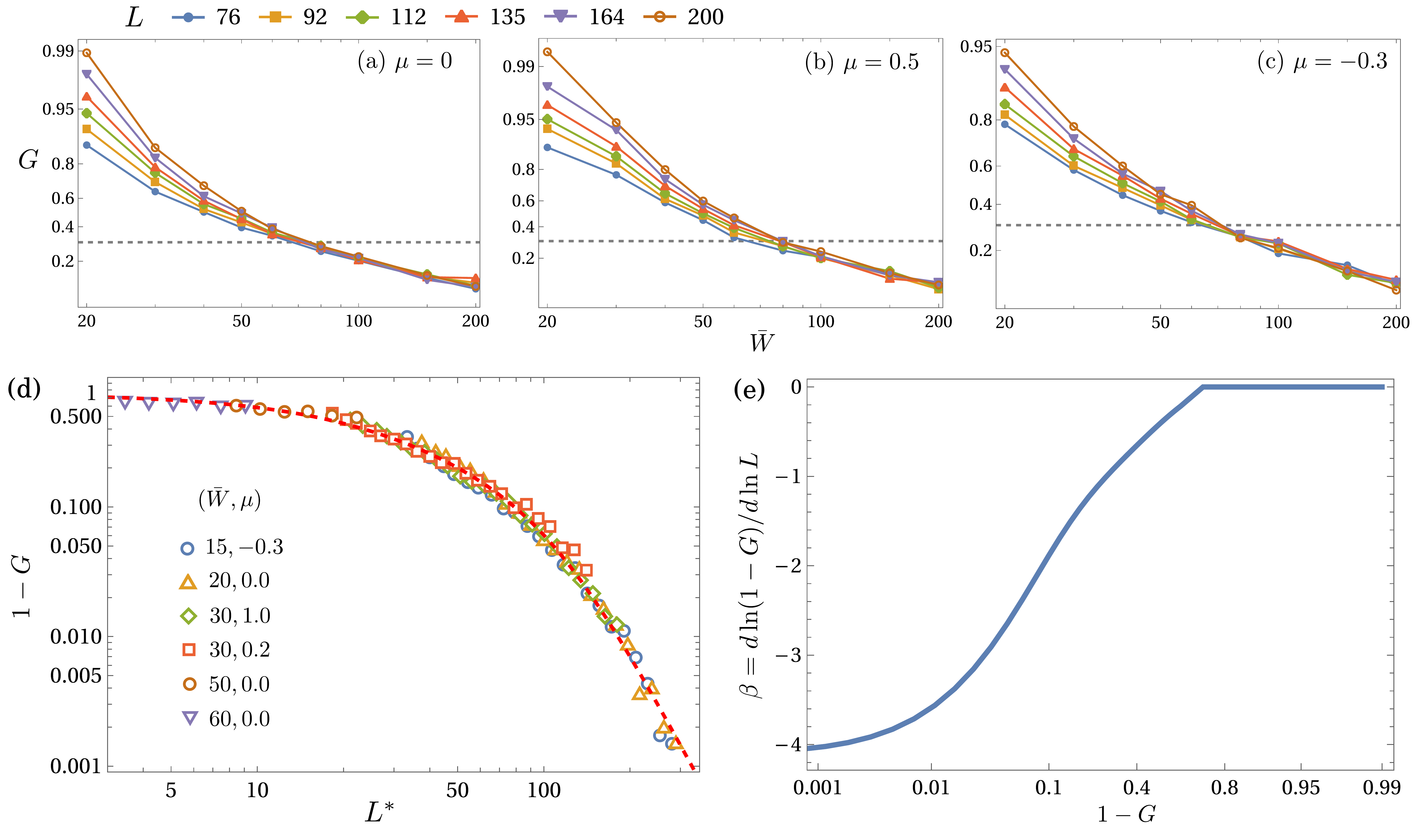}
	\caption{(a-c) Finite-size scaling across the WQTI-AI transition with respect to the disorder strength $\bar{W}$ at different applied chemical potential $\mu$. (d) Quantization error with respect to the scaled size. The red dashed line is fitted from the numerical data. (e) Scaling exponent derived from the fitted line in (d). The $1-G$ axis in (e) is plotted in the $\ln[x/(1-x)]$ scale. The lattice parameters are similar to Fig.~(1) in the main text.  \label{fig3}}
\end{figure}

In this subsection, we provide more numerical evidence to support our one-parameter scaling ansatz. In Figs.~\ref{fig3}(a-c), we present the scaling trend across the WQTI-AI transition at different chemical potentials $\mu$. Within the error caused by random fluctuations, there is little size dependence for $G<0.3$, supporting that the WQTI-AI critical value is universal given a certain set of lattice parameters. In the main text, we already demonstrate the one-parameter scaling function within the WQTI phase. Now to fully map out this function, we add two more sets of data near the WQTI-AI phase boundary as shown in Fig.~\ref{fig3}(d) and are able fit the collapsed data in a smooth line (dashed line). By taking the log derivative of the fitted line, we compute the scaling exponent $\beta=d\ln(1-G)/\ln L$ in Fig.~\ref{fig3}(e). The scaling exponent approaches $-4$ in the limit $G\to 1$, then graduate increases until the critical value $G\sim 0.3$ and stays zero afterward. For the transverse conductance, one should see the $\beta$ function cutting through zero axis at $G=0.5$ and turning positive afterward. We explain the reason in the main text.

\subsection{Other types of disorders}
\begin{figure}
	\includegraphics[width=0.7\textwidth]{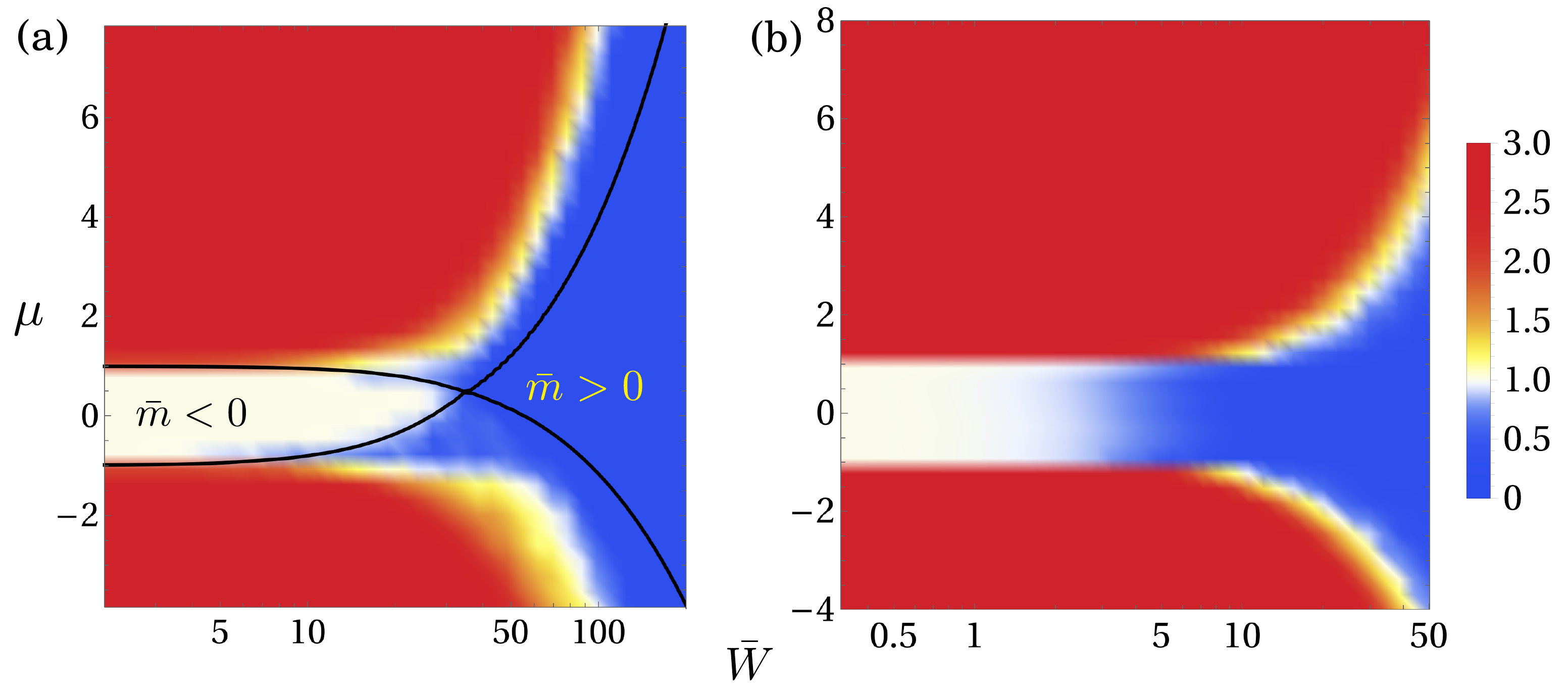}
	\caption{Conductance in the presence of SOC (a) and magnetic (b) disorders. The black lines in panel (a) represent the renormalized energy gap with the mass term inverted. The parameters for the clean system are similar to Fig.~2(a) in the main text with $\xi=2$. For the symmetry-breaking disorder, the quantized conductance is lost at $\bar{W}\sim \mathcal{O}(1)$; while in the presence of symmetry-preserving disorder the quantization plateau can extend up to $\bar{W}\sim \mathcal{O}(10^2)$\label{fig4}. }
\end{figure}

In this subsection, we consider the full time-reversal invariant Hamiltonian, i.e. two copies of the Hamiltonian we used in the main text, so that $H_{AII}(\kk)=h(\kk)\oplus -h^*(-\kk)$ with the time-reversing operator $i\mathcal{K}\tau_y$ and $\mathcal{K}$ being the complex conjugate. Here $\tau_{x,y,z,0}$ are the Pauli matrices corresponding to the spin degrees of freedom. Two more types of disorders can be studied in this case, i.e. the time-reversal symmetric spin-orbit coupling (SOC) disorder $u(\rr)\tau_x\sigma_y$ and the time-reversal breaking magnetic disorder $u(\rr)\tau_x\sigma_x$. In the former case, the difference from the onsite potential studied in the main text is that in one spin sector
\begin{equation}
	\Sigma = \sigma_y\left[\frac{W^2a^2}{12}\int \frac{d^2\kk}{(2\pi)^2} \frac{e^{-k^2\xi^2/2}}{\mu-h(\kk)-\Sigma}\right]\sigma_y.
\end{equation}
Since $\sigma_y\sigma_z\sigma_y=-\sigma_z$, the sign of the mass correction is inverted. As a result, a clean non-trivial system ($m<0$) will be eventually trivialized when $\bar{m}>0$. This is reflected in Fig.~\ref{fig4}(a). On the other hand, a clean trivial system $(m>0)$ continues to stay trivial. For the case of time-reversal symmetry breaking disorder, the quantization plateau is destroyed at a much weaker strength of disorder compared to the cases of symmetry-preserving disorders [see Fig.~\ref{fig4}(b)], consistent with the fact that the topology of AII systems is protected by the time-reversal symmetry.

\subsection{Pseudo-quantized conductance plateau}
\begin{figure}
	\includegraphics[width=0.5\textwidth]{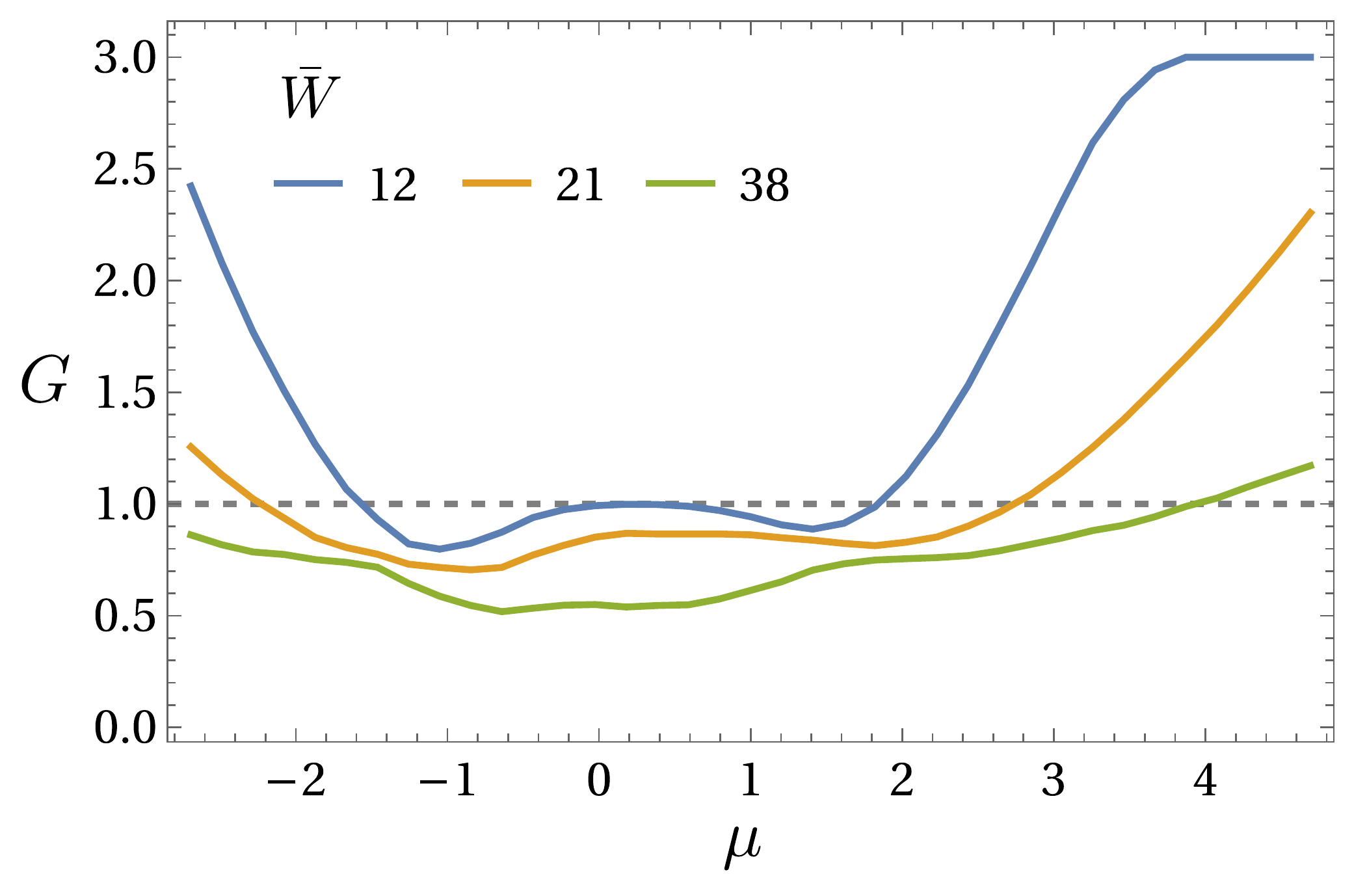}
	\caption{Conductance as a function of the applied chemical potential for different strengths of disorder. The parameters for the clean system are similar to Fig.~2(c) in the main text with $\xi=8$. \label{fig5}. The quantization error of the conductance plateau gets worse as the disorder strength increases, driving the system from TI to WQTI phase.}
\end{figure}

In Fig.~\ref{fig5}, we show that in the WQTI phase, due to the slow convergence of the conductance to the quantized value (especially near the WQTI-AI transition), it is possible to observe a conductance plateau by tuning the chemical potential but the plateau value is not precisely quantized.

